# The Open-Weight Paradox:
# Why Restricting Access to AI Models May Undermine the Safety It Seeks to Protect

Vinicius Santana Gomes[1]

**Abstract**

The governance of open-weight artificial intelligence (AI) models has been framed as a binary choice: openness as risk, restriction as safety. This paper challenges that framing, arguing that access restrictions, without governed alternatives, may displace risks rather than reduce them. The global concentration of compute infrastructure makes open-weight models one of the most viable pathways to sovereign AI capacity in the Global South; restricting such access deepens asymmetries while driving proliferation into unsupervised settings. This analysis proposes that hardware-layer governance, including chip-level attestation mechanisms such as FlexHEG, trusted execution environments, confidential computing, and complementary software-layer safeguards, offers a defense-in-depth alternative to the current binary. A threat model taxonomy mapping misuse vectors to hardware, software, institutional, and liability layers illustrates why no single governance mechanism suffices. To operationalize this approach, the paper argues that effective AI governance as a dual-use technology will likely require a multilateral institutional architecture functionally analogous, though not identical, to the role performed by the IAEA in the nuclear domain, with explicit safeguards against the co-option of hardware controls for domestic repression. The relevant policy question is how to make openness safer through technical and institutional design while addressing the transition realities of legacy hardware, attestation at scale, and civil liberties protection.

**Keywords:** open-weight AI; AI governance; hardware-enabled governance; FlexHEG; IAEA; AI safety; AI sovereignty; compute infrastructure.

## I. Introduction

The dominant policy debate over open-weight AI models has consolidated around a binary framing: open equals dangerous, closed equals safe. Proponents of restriction argue that releasing model weights allows malicious actors to strip safety guardrails, weaponize those model capabilities, and evade accountability. Proponents of openness counter that transparency accelerates research, distributes power, and prevents monopolistic concentration. Both positions have merit, but neither adequately describes the risks as they actually present themselves.

Throughout this text, "open-weight" refers to making a model's trained weights available for download, execution, and adaptation by third parties, which is not necessarily synonymous with fully open-source software, nor with mere remote API access. This distinction matters because the core regulatory problem discussed here is not merely code transparency, but the concrete possibility of local execution, fine-tuning, and deployment without ongoing dependence on a centralized provider.

This paper advances a different thesis: the security risks most worth examining may not lie in openness itself, but in what occurs when access is restricted without governed alternatives. When the primary pathways to advanced AI capability run through a handful of cloud providers controlled by a small number of nations and corporations, the resulting arrangement tends to amplify structural dependency and opacity, without by itself guaranteeing greater safety.

The release of DeepSeek R1 in January 2025 made this dynamic concrete. A Chinese AI company, operating under severe hardware export restrictions, produced an open-weight reasoning model that claimed to rival the performance of leading proprietary systems at a fraction of the cost (Nandi & Balasubramani, 2025). The episode exposed two dynamics: export controls and access restrictions alone are insufficient to prevent the emergence of highly capable models; and open-weight architectures can function as instruments of capability diffusion for actors excluded from the frontier compute ecosystem. As Quansah (2025) observed in analysis for the Carnegie Endowment, DeepSeek R1 weakened the assumption that advanced AI would remain accessible only to wealthy nations and large multinational corporations.

---

[1] Special Advisor, Government of the Federal District of Brazil; AI Instructor, School of Government of the Federal District of Brazil. Contact: vinicius.gomes@sedet.df.gov.br; gomes.vs@gmail.com

The question, therefore, is not whether to permit open-weight models. They are already, in part, available, and transnational diffusion and digital copiability make it unlikely that unilateral restrictions can sufficiently contain their proliferation from any single jurisdiction. The question is how to build a technical and institutional infrastructure that can reduce the risks of openness through verifiable safeguards.

Under current institutional and geopolitical conditions, policies focused solely on restricting access to open-weight models may reduce visibility without proportionally reducing risk. The analysis examines this problem at three levels: infrastructure, the technical vulnerability of fine-tuning, and the institutional design of governance.

## II. Infrastructure, Dependency, and Risk Displacement

The vast majority of nations and organizations cannot train or host frontier AI models. Only a few countries possess the compute infrastructure, energy supply, and technical talent required to develop systems at the technological frontier.

In most nations, particularly outside the dominant compute axis, data centers lack GPUs for training or advanced inference. The local choice typically falls between running limited models on conventional servers or routing data to external providers via API, with implications for data sovereignty and data protection, precisely the dependency that restrictive policies would deepen.

When access is severely restricted without governed alternatives, three consequences become likely: proliferation of Shadow AI through informal channels, compounded by the market concentration identified as a systemic risk (Bengio et al., 2025); deepening dependence on foreign cloud infrastructure; and migration of demand to providers outside Western regulatory frameworks. The result is that activity migrates from partially auditable environments to ones where oversight is minimal.

Emerging evidence from the post-2022 export control regime lends empirical support to this displacement pattern. On the supply side, Sastry et al. (2024) observe that "actors are likely to attempt to circumvent and evade compute governance interventions, especially where their access to AI chips or their privacy is severely affected," noting that in the short term, circumvention occurs "via chip smuggling, using non-controlled chips, or accessing cloud compute," while in the medium term, "squeezing one part of the supply chain puts pressure on other parts" and "actors without access to high-end chips are incentivized to find ways to utilize larger quantities of lower-grade chips". Aarne, Fist, and Withers (2024) confirm that "AI chip smuggling is already happening today and could significantly grow in volume over the coming years". By December 2025, the U.S. Department of Justice had dismantled a network that between October 2024 and May 2025 smuggled at least $160 million worth of export-controlled Nvidia H100 and H200 GPUs, with falsified shipping paperwork concealing the ultimate destination of the chips (U.S. DOJ, 2025). On the demand side, the open-weight model ecosystem has expanded at a pace that complicates enforcement: by 2025, Hugging Face hosted more than two million public models, with users increasingly creating derivative artifacts such as fine-tuned models, adapters, benchmarks, and applications rather than only consuming pre-trained systems (Ghosh et al., 2026). Local deployment tools such as Ollama and LM Studio have made it possible to run quantized versions of these models on consumer hardware, effectively decoupling inference from cloud infrastructure and from the monitoring that cloud environments afford. At the enterprise level, a Gartner

survey of 302 cybersecurity leaders conducted between March and May 2025 found that 69% of organizations suspected or had evidence that employees were using prohibited public generative AI tools, a phenomenon Gartner projects will produce security or compliance incidents in more than 40% of enterprises by 2030 (Gartner, 2025).

These developments suggest a tripartite displacement dynamic that varies with the policy regime in place. Under restriction without governed alternatives, the scenario examined here, demand migrates from auditable channels to shadow infrastructure, offshore providers, and illicit supply chains, reducing rather than increasing the overall legibility of the ecosystem. Under restriction with governed alternatives, such as monitored cloud environments paired with subsidized access, a larger share of activity can be retained within frameworks where auditing is feasible, though residual displacement persists for actors specifically seeking to evade oversight. Under openness with hardware-layer governance, the approach this paper argues for, verifiable safeguards operate at the substrate level, permitting wide distribution of model weights while maintaining compliance monitoring at the point of compute.

Sovereign AI development requires access to high-performance GPUs, hyperscale data centers, and energy capacity to sustain large training runs, resources that most Global South countries do not possess. Without this specialized hardware, neither capacity-building investments, nor regulatory reforms, nor national AI strategies can, on their own, enable training proprietary models. Compute infrastructure comes before other barriers because without it, large-scale training is not practically feasible. The 2024 update to the OECD AI Principles recognized this dimension, emphasizing investment in open-source tools with attention to emerging economies (OECD, 2024).

The empirical evidence is robust. Lehdonvirta, Wú, and Hawkins (2024) found that only 39 countries host public cloud regions, of which 30 offer GPUs, with China and the United States together hosting nearly as many GPU-enabled regions (49) as the rest of the world combined (52). The authors propose three categories: "Compute North" (chips for training), "Compute South" (chips for inference), and "Compute Desert" (no GPUs available). Data from Epoch AI (Pilz et al., 2025) confirm: the United States concentrates roughly three-quarters of global GPU cluster performance, with China in second place at approximately 15%.

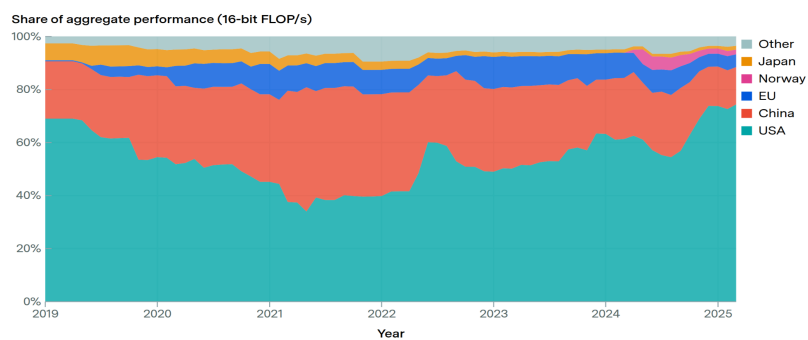

**Figure 1.** Share of aggregate AI supercomputer performance (16-bit FLOP/s) by country, 2019–2025. Source: Epoch AI (2025). Used under Creative Commons BY license.

This impossibility of training proprietary models is what makes open-weight models one of the primary mechanisms for expanding AI capacity accessible to the Global South. Because the model weights have already been trained by third parties, nations and organizations without GPUs can download smaller versions of these models, sized to run on conventional servers and available hardware, and adapt them locally to their needs, partially bypassing a hardware barrier that is otherwise insurmountable.

The problem is that the same open weights that enable sovereignty also permit the removal of safety guardrails and unsupervised deployment. The equalizing potential of open-weight models and the security risks that accompany them are inseparable.

A plausible objection to this framing is that the compute divide may be transitory, following the typical diffusion pattern of capital-intensive technologies: initial concentration among a few producers and early adopters, followed by progressive commoditization. The empirical trajectory of inference costs lends partial support to this view. According to Stanford's 2025 AI Index Report, the inference cost for a system performing at the level of GPT-3.5 fell over 280-fold between November 2022 and October 2024, with hardware costs declining approximately 30% annually and energy efficiency improving by 40% per year (Maslej et al., 2025). Epoch AI's analysis of inference pricing confirms that, across multiple benchmarks, prices have been declining at rates ranging from 9-fold to 900-fold per year depending on the performance milestone, with mid-range benchmarks showing declines of approximately 40-fold per year (Cottier et al., 2025).

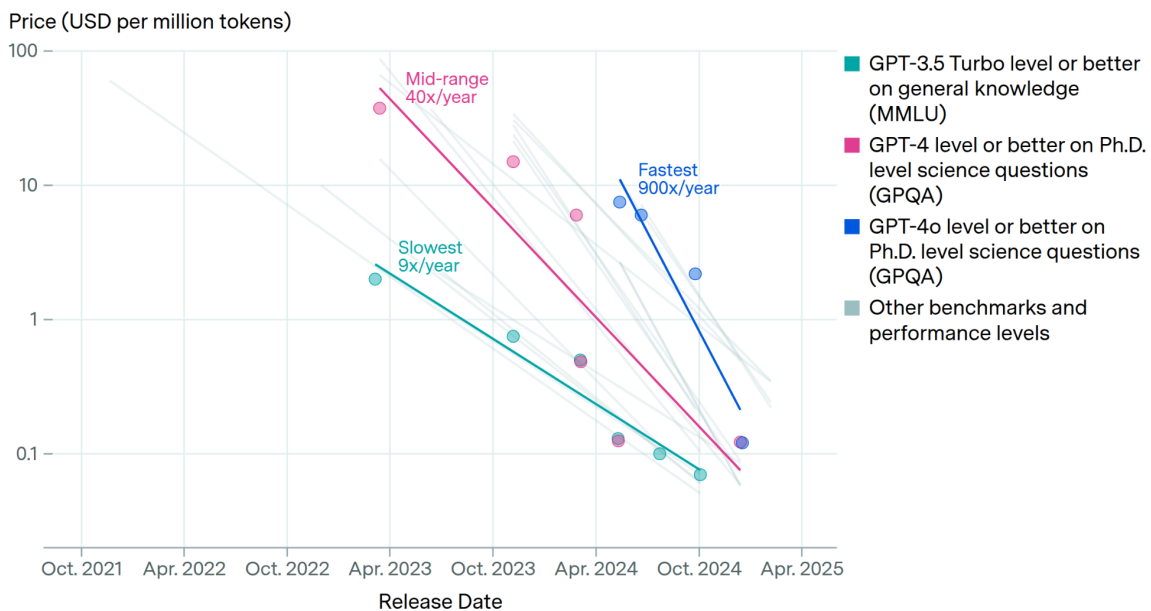

**Figure 2.** LLM inference price across select benchmarks and performance levels, 2021–2025. Source: Cottier, B., Snodin, B., Owen, D., & Adamczewski, T. (2025). Epoch AI. Used under Creative Commons BY license.

These trends are significant, but the commoditization thesis requires important qualifications. First, the most dramatic cost reductions have occurred for inference on already-trained models, not for the training process itself, which remains the primary problem for sovereign capability. A nation that can deploy a third-party model cheaply via API or local inference has not thereby acquired the capacity to train, adapt, or align models to its own institutional, linguistic, or regulatory context. Second, even as inference costs fall, the compute infrastructure required for advanced training runs continues to concentrate. Third, the commoditization dynamic does not eliminate the dependency problem; it restructures it. If inference becomes cheap through commercial APIs, the structural dependency shifts from hardware access to platform dependency, routing national data through foreign providers with implications for sovereignty that persist regardless of unit cost. Fourth, the very decline in inference costs accelerates the risk dynamics examined in this paper: cheaper and more accessible models amplify the fine-tuning vulnerability and the proliferation of unsupervised deployment, making governance mechanisms more urgent.

Beyond this structural dependency, a further challenge deserves attention. In February 2026, the Qwen3.5 family made visible a trend already underway in the open-weight ecosystem: the public circulation of smaller models accompanied by official deployment documentation and quantized variants for local execution. The case matters less for its specific benchmark than for what it illustrates: the assumption that only large actors with powerful compute would have meaningful access to open-weight models is eroding (Qwen Team, 2026; MLX Community, 2026). Models that are locally executable and potentially modifiable substantially amplify the risk landscape.

Moreover, these locally executable models may fall below the thresholds that trigger the most substantive obligations under regulations such as the EU AI Act (Regulation (EU) 2024/1689).

Existing regulatory approaches do not fully resolve this dilemma. The EU AI Act presumes that models trained above $10^{25}$ floating-point operations have high-impact capabilities, which triggers their classification as GPAI[2] models with systemic risk (Article 51(1)(a) and 51(2), Regulation (EU) 2024/1689). The Commission's implementing Guidelines further introduce $10^{23}$ floating-point operations, combined with the ability to generate language, images, or video from text, as an indicative criterion for classifying a model as general-purpose AI (European Commission, 2025).

Even where such models may technically qualify as GPAI under the indicative $10^{23}$ floating-point operations criterion, the modifications that pose the greatest safety risks require negligible computational resources, a regulatory gap examined in detail in Section III.

This does not diminish the importance of the AI Act as a relevant regulatory advance. The point is narrower: criteria based predominantly on training scale are useful for part of the problem, but leave less visible the dynamics of adaptation, redistribution, and decentralized deployment of models below those thresholds.

---

[2] Article 51(2), Regulation (EU) 2024/1689, establishes a rebuttable presumption of systemic risk when training compute exceeds $10^{25}$ FLOP. The $10^{23}$ FLOP indicative criterion for GPAI classification derives from the European Commission's non-binding guidelines of July 2025: models exceeding this threshold that can generate language, text-to-image, or text-to-video are presumptively considered GPAI models, though this is not an absolute rule. See: European Commission, Guidelines on the scope of obligations for providers of general-purpose AI models under the AI Act (2025).

This design illustrates a practical limitation of compute-anchored regulations: they organize obligations well for those who train the most advanced models, but have fewer levers over the ecosystem of smaller open-weight models that circulate as downloadable weights and can be adapted and deployed on consumer hardware, a common scenario outside major compute hubs, although some of these uses may still be captured by rules applicable to the final system and its context of use.

The risk is that restriction policies designed for jurisdictions with abundant resources deepen, in others, dependencies on foreign infrastructure and regulation rather than increasing safety.

**III. The Fine-Tuning Vulnerability: Why Restriction Misses the Target**

The main argument for restricting open-weight models is that releasing model weights allows malicious actors to remove safety guardrails through fine-tuning. This concern has empirical grounding. Qi et al. (2024) demonstrated that safety alignment can be compromised by fine-tuning with as few as ten adversarially designed examples, at a cost of less than $0.20 via OpenAI's APIs. The UK AI Security Institute confirmed that current anti-tampering techniques remain insufficient: safeguards designed to prevent harmful behaviors can be removed with just a few dozen training examples, in a matter of minutes (UK AISI, 2025).

However, this vulnerability is not exclusive to open-weight models. Qi et al. (2024) demonstrated the attack on both open-weight models (Llama-2) and closed API models (GPT-3.5 Turbo), showing that the fine-tuning risk extends to any model for which fine-tuning access is provided, which includes most commercial offerings. Existing regulation also fails to address this vulnerability. The GPAI Guidelines (European Commission, 2025) establish that only significant modification to models, not minor adjustments, trigger provider obligations for the modifier. But the most dangerous modifications, those that remove safety guardrails, require negligible computational resources, falling below any significant modification threshold. This creates a regulatory blind spot: the modifications that matter most fall below the line of regulatory visibility.

The implication: restricting the distribution of open-weight models addresses one vector of the fine-tuning problem while leaving the more fundamental vulnerability untouched. A policy framework that restricts distribution without addressing the fine-tuning mechanism itself leaves the most vulnerable technical core of the problem intact.

The regulatory impasse stems less from normative disagreement about openness than from a mismatch between the object being controlled and the technical layer at which control would need to operate.

If the decisive vulnerability lies not only in the distribution of weights, but in the material possibility of reconfiguring and running models outside verifiable oversight mechanisms, then the most promising layer for intervention shifts from software circulation to the compute infrastructure that underpins training and execution.

The fine-tuning vulnerability, however, is one vector within a broader threat landscape. Table 1 maps five principal threats associated with open-weight AI models to the governance layer best positioned to mitigate each.

Table 1. Threat model taxonomy: governance layers and mitigation mechanisms

| Threat Vector | Hardware Layer | Software Layer | Institutional Layer | Liability Layer |
|---|---|---|---|---|
| **Safety guardrail removal via fine-tuning** | Attestation logging of training runs; compute caps on unattested workloads (Petrie et al., 2025) | Pre-release safety evaluations; alignment re-testing after modification (UK AISI, 2025) | Red-team mandates; model-release norms | Product liability for deployers of modified models (Directive (EU) 2024/2853) |
| **Model extraction / distillation** | TEE-based weight protection; confidential inference (Mohanty et al., 2025) | Watermarking; model fingerprinting | Provenance registries; IP licensing frameworks | Intellectual property liability for unauthorized extraction |
| **Covert large-scale deployment** | Chip-level attestation; single-instance enforcement (Vasileiadis et al., 2025) | API metering; runtime monitoring | Audit obligations; reporting requirements | Platform liability for hosting unattested deployments |
| **Sanctions evasion** | Geofencing; export verification via on-chip mechanisms (Kulp et al., 2024; Aarne, Fist, and Withers, 2024) | — | Export control enforcement; multilateral compliance frameworks | Export control penalties for suppliers and intermediaries participating in diversion; liability for compute providers with insufficient due diligence (Kulp et al., 2024; Sastry et al., 2024) |
| **Scaled misuse (CBRN, cyber)** | Compute caps on unattested workloads | Output filtering; use-case restrictions | International safety standards; incident-reporting obligations | Criminal and civil liability for foreseeable harms |

Two features of the governance problem are visible in this mapping. First, no single layer addresses every threat vector. Hardware mechanisms cover execution-level risks but do not themselves create institutional accountability; liability mechanisms create incentives for commercial actors but leave non-commercial misuse unaddressed. Effective governance requires all four layers in coordination. Second, the liability column identifies where legal responsibility binds most concretely: deployers of modified models bear product liability under the EU's revised Product Liability Directive; platforms hosting unattested deployments face platform liability; and actors causing foreseeable harm through scaled misuse are subject to criminal and civil liability. These legal mechanisms create economic incentives that technical safeguards, on their own, do not generate, a point developed further in Section VI.

Software-layer mechanisms complement this architecture. Model-release norms, including pre-release safety evaluations, red-teaming, and staged deployment, provide an initial filter against the most foreseeable harms; the International AI Safety Report documented their growing adoption alongside the finding that companies publishing Frontier AI Safety Frameworks more than doubled during 2025 (Bengio et al., 2025b). Watermarking and provenance tracking provide post-hoc attribution: they cannot prevent misuse, but they reduce the practical anonymity of model-derived outputs by connecting generated content to its source. These mechanisms are limited by a constraint documented throughout this section: safety guardrails at the model level can be removed through fine-tuning with minimal computational resources (Qi et al., 2024; UK AISI, 2025). Watermarks face analogous robustness limitations, as the same report documented that sophisticated attackers frequently manage to circumvent current software-level defenses (Bengio et al., 2025b; UK AISI, 2025). Their value is not standalone sufficiency but their contribution to the layered architecture that the table makes explicit.

## IV. Hardware-Layer Governance: From FlexHEG to a Landscape of Verifiable Compute

### A. The Case for Hardware-Layer Governance

If software-level restrictions are insufficient and distribution restrictions are counterproductive, where should governance be applied? Several recent proposals point to the hardware layer, exploring the possibility of embedding safeguards directly into compute infrastructure. These proposals do not yet offer a consolidated solution, but they point in a useful direction: governance that operates over the verifiable conditions of execution, not just software distribution.

The FlexHEG (Flexible Hardware-Enabled Guarantees) initiative proposes adding security systems to AI chips that combine a tamper-proof enclosure with a guarantee processor, enabling privacy-preserving verification and enforcement of compliance over how chips are used (Petrie et al., 2025). These mechanisms could be used flexibly to solve coordination problems as they arise, allowing states and other actors to make credible and verifiable commitments.

The approach preserves privacy because verifications can be aggregated across devices to produce abstract conclusions, such as "this model was trained with less than X FLOP," without revealing technical details. Rules are programmable and updatable, which avoids the accumulation of obsolete regulations. Rule configuration can also be assigned to different actors, from the chip owner to a quorum of states, enabling multilateral governance without a single centralized authority. The design would be open-source, auditable, and robust against tampering even by state-level adversaries (Petrie et al., 2025).

These advantages, however, are conditional on significant challenges. Full FlexHEG mechanisms would primarily be incorporated into new chips, though partial capabilities might be retrofitted onto existing hardware; legacy chips would continue operating for years with limited or no guarantees, creating a transition period with partial coverage. The frontier chip manufacturing chain is highly concentrated, essentially Nvidia in design and TSMC in fabrication, which implies dependence on cooperation from very few private actors. The FlexHEG report itself acknowledges that early versions may need to compromise between security and the sophistication of supported rules (Petrie et al., 2025).

The aviation analogy is useful in a strictly functional sense: high-risk systems tend to be governed through technical certification, continuous monitoring, failure investigation, and standard updates — not merely through prohibition or unrestricted release. FlexHEG transposes this logic to compute infrastructure: safety is embedded in the hardware, not dependent on restricting who can use it.

The aviation analogy, however, should be extended to a broader principle: defense in depth. In safety-critical domains, effective governance rarely depends on a single mechanism. Regulatory regimes for aviation, nuclear energy, and medical devices share a common structural feature: they layer multiple independent safeguards so that the failure of any single mechanism does not produce catastrophic outcomes. The FDA's regulation of software as a medical device, for instance, establishes a layered regulatory framework that integrates risk management throughout the entire product lifecycle. Premarket guidance recommends that sponsors include documented risk assessment, verification and validation processes, and cybersecurity considerations in their submissions (FDA, 2023), while the Quality Management System Regulation requires manufacturers to maintain risk-based quality management systems encompassing design controls, production processes, complaint handling, corrective and preventive actions, and post-market reporting obligations (FDA, 2024).

The International AI Safety Report's second key update similarly documented that current risk mitigation methods can be circumvented by sophisticated actors, that safeguards vary in effectiveness across deployment contexts, and that developers are increasingly adopting a defense-in-depth strategy that layers multiple safeguards across the AI lifecycle, from training interventions through deployment controls to post-deployment monitoring, so that the failure of any single layer does not produce harmful outcomes (Bengio et al., 2025b). The report also documented that governments and international organizations have begun translating technical safeguards into governance requirements, embedding model evaluations, provenance tracking, and red-teaming within broader oversight structures (Bengio et al., 2025b).

FlexHEG should therefore be understood not as a standalone solution, but as one critical layer within a defense-in-depth architecture for AI governance. Hardware-level verification addresses the compute substrate; software-level safeguards address model behavior; institutional oversight addresses accountability and legitimacy. No single layer is sufficient. Without hardware-level verification, the other layers rest on an unverifiable foundation.

**B. The Broader Landscape: TEEs, Confidential Computing, and Verifiable Training**

FlexHEG is not the only mechanism through which hardware can anchor governance. Trusted Execution Environments, confidential computing platforms, and cryptographic verification schemes already provide, or are close to providing, several of the governance functions this paper attributes to the hardware layer. FlexHEG is better understood as one point on a spectrum of enforceable mechanisms, not as a singular proposal awaiting implementation from scratch.

Hardware attestation can do more than verify enclave identity at launch. TALOS, a framework for verifiable state management during enclave migration, introduces a Proof of Execution (PoX) mechanism that "monitors control-flow transitions in real time to detect anomalies such as code injection, unauthorized forks, or unexpected system calls" and "continuously verifies that execution adheres to the expected control-flow graph and originates from an unmodified, approved binary" throughout the migration cycle (Vasileiadis et al., 2025). TALOS also enforces "an one-to-one mapping between application instances and their system-level identifiers, preventing replication and fork-based cloning attacks," which is relevant to single-instance semantics for governed AI workloads. The framework has been prototyped on both Intel SGX and RISC-V Keystone, and "its design is agnostic to the underlying TEE architecture" (Vasileiadis et al., 2025).

On the lifecycle management side, Keyfort extends RISC-V Keystone TEEs with modules for "secure enclave-local timekeeping, state continuity, secure software updates and migrations, allowing enclaves to evolve securely across software versions and hardware devices" (Wilde et al., 2026). These additions, covering trusted time, anti-rollback protection, and secure migration, add approximately 750 lines of code to the trusted computing base and limit service downtime to 10 ms for updates and 0.7 s for migrations of enclaves with up to 16 KB of state (Wilde et al., 2026). State continuity and anti-cloning guarantees matter for AI governance because they prevent covert model replication and ensure that policy-compliant lifecycle controls survive hardware transitions.

Full chip-native governance is not yet available for most hardware. Hybrid approaches can nonetheless protect model weights against extraction and unauthorized distillation. DistilLock runs "the proprietary foundation model as a black-box teacher inside a hardware-protected enclave on the user's device," allowing knowledge distillation to proceed on-device "without exposing the model weights or user data" (Mohanty et al., 2025). To manage performance costs, DistilLock "employs a model obfuscation strategy that enables most compute-heavy operations to be offloaded to untrusted accelerators (e.g., GPUs)," with the TEE handling only lightweight authorization (Mohanty et al., 2025). TBNet takes a different approach, generating a "Two-Branch substitution model" that places a secure branch inside the TEE and an unsecured branch in the Rich Execution Environment, enforcing "layer-wise connections from REE to TEE, but not vice versa" (Liu et al., 2024). This one-way information flow is designed to protect the confidentiality of TEE-protected layers and reduces TEE memory usage by up to 2.45× compared to full in-enclave execution (Liu et al., 2024). These frameworks show that weight protection is achievable on current hardware, without purpose-built governance chips.

The practical reach of TEE-based governance is limited by the fact that most confidential computing implementations remain CPU-bound. Dong and Wang (2025) present the first performance evaluation of the DeepSeek model within a TEE-based confidential computing environment using Intel TDX. TDX outperforms standard CPU execution for smaller models: DeepSeek-R1-1.5B achieved 25.67 tokens/s under TDX versus 10.25 tokens/s on CPU alone. But "the overall GPU-to-CPU performance ratio averages 12 across different model sizes," and "current confidential computing solutions primarily rely on CPU-based TEEs with limited resources, restricting their ability to support large models like LLMs." A companion study confirms that "the communication between the CPU host and the GPU device currently occurs in plaintext, lacking adequate safeguards for protecting confidential data associated with SoC design processes" (Dong, Feng, and Wang, 2025). TEE-based governance is therefore viable for smaller and distilled models now, but extending it to frontier-scale systems requires closing the GPU trust boundary, a problem that both NVIDIA's confidential computing extensions and the FlexHEG proposal target through different technical strategies.

Hardware attestation is most effective when combined with institutional and cryptographic mechanisms at other layers. Sastry et al. (2024) argue that compute is a "particularly effective point of intervention" because "it is detectable, excludable, and quantifiable, and is produced via an extremely concentrated supply chain." They propose that governance of compute can enhance three capacities: increasing regulatory visibility into AI capabilities and use, allocating resources toward safe and beneficial uses of AI, and enforcing prohibitions against reckless or malicious development or use. Compute reporting, monitoring, and subsidized access provide institutional mechanisms within which hardware attestation could operate.

Zero-knowledge proofs offer a separate pathway: verifying computational claims without revealing model internals. A comprehensive survey of ZKML research from 2017 to 2024 categorizes existing work under "three key categories: verifiable training, verifiable inference, and verifiable testing," and concludes that "ZKP ensures the verifiability and security of machine learning models, making it a valuable tool for privacy-preserving AI" (Peng et al., 2025). Laminator takes this further by proposing "the first framework for verifiable ML property cards via hardware-assisted ML property attestations," allowing model providers to prove properties such as accuracy and fairness and, when combined with signed external certificates from trusted entities, draw conclusions about dataset provenance; its attestations are "efficient in terms of overhead, scalable to large numbers of verifiers, and versatile with respect to the properties it can prove during training or inference" (Duddu et al., 2025). Such attestations could form the technical basis for compliance verification under the multilateral oversight architecture proposed in Section V.B.

These mechanisms are not competitors; they occupy different layers of a defense-in-depth architecture. Hou, Zhao, and Wang (2025) propose a three-layer framework for the next generation of LLM application ecosystems: an infrastructure layer that "combines models, computing, networks, and data as the base support for LLMs" and "decides how large a model can run, how fast it works, and how safely it interacts"; a protocol layer that "defines standards for communication and coordination across agents, services, and devices"; and an application layer that "offers the user-facing intelligence." Governance maps onto this structure naturally. Hardware attestation and TEE-based controls operate at the infrastructure layer; communication and coordination standards shape the protocol layer; safety evaluations and release norms operate at the application layer. The policy question is how to compose these layers so that each covers the gaps left by the others, rather than searching for a single mechanism capable of bearing the full weight of enforcement alone.

**C. Implementation Challenges and Transition Realities**

Available mechanisms still leave practical problems unaddressed. Three problems require direct treatment: the transition from ungoverned legacy hardware, the problem of offline and air-gapped deployments, and the definition of a minimal near-term rule set that avoids both overreach and under-specification.

Data center AI chips already in circulation, including chips that predate any attestation capability, will remain operational for a substantial transition period even if new generations are required to incorporate governance mechanisms. During this transition, a significant fraction of global compute capacity will exist outside any hardware governance regime. This is a genuine limitation, but not a fatal one. The Montreal Protocol's elimination of chlorofluorocarbons and the phaseout of leaded gasoline both tolerated multi-year transition windows while mandating that new production meet stricter standards. Governance need not achieve universal coverage to be effective; it must govern the margin of new compute entering the ecosystem. If all newly manufactured AI accelerators ship with attestation capability and are subject to governance protocols, the fraction of ungoverned compute will decline as governable hardware diffuses through the installed base. Chip smuggling and diversion remain residual risks, but they can be mitigated through export control enforcement and chip ownership or supply-chain tracking mechanisms, together with inspections where required (Kulp et al., 2024; Aarne, Fist, and Withers, 2024).

Not all AI deployments maintain persistent network connectivity. Military systems, industrial control environments, and sovereignty-sensitive government installations may operate in air-gapped or intermittently connected configurations. A governance framework that depends entirely on real-time attestation would be inapplicable in these contexts, a point the reviewer rightly raises. Hardware governance can, however, accommodate offline operation through several mechanisms. Hardware operating licenses can be verified locally, entirely offline, and renewed periodically when connectivity or physical delivery is available; secure modules can also support remote attestation and, where implemented, secure logging in non-volatile memory; and license expiration can trigger throttling or shut-off rather than requiring constant online connectivity. Rather than assuming continuous real-time attestation, the governance framework can rely on periodic reauthorization, remote attestation when connectivity is available, and inspections where required. These approaches sacrifice some real-time enforcement capacity in exchange for preserving sovereignty and operational flexibility, a tradeoff consistent with existing practice in nuclear safeguards, where physical inspections complement continuous monitoring.

The most consequential design decision is what the initial rule set should enforce. Over-specification risks false positives and compliance costs that push users toward ungoverned alternatives. Too little specificity, on the other hand, makes the framework irrelevant. This paper proposes three minimal rules as a starting configuration for FlexHEG-capable and TEE-attested hardware:

(1) Training-run attestation logging. All training runs exceeding a defined compute threshold (calibrated to remain well below frontier capability levels) must generate a cryptographically signed attestation record of the relevant metered quantities and the hardware identifiers involved. This creates a verifiable record without constraining what can be trained (Kulp et al., 2024; Aarne, Fist, and Withers, 2024).

(2) Single-instance semantics for high-capability models. Models whose training compute exceeds a higher threshold, corresponding roughly to current frontier models, must be bound to attested deployment instances, preventing unauthorized replication. The anti-cloning and state-continuity mechanisms demonstrated by TALOS and Keyfort (discussed in Section IV.B) provide part of the technical foundation for this kind of requirement (Vasileiadis et al., 2025; Wilde et al., 2026).

(3) Periodic compliance check-ins. Hardware operating under governance protocols must attest its compliance status at regular intervals, with the attestation verified by the relevant licensing or oversight authority. Failure to check in within the specified window triggers capability restrictions, not immediate shutdown, preserving operational continuity while incentivizing compliance (Kulp et al., 2024; Aarne, Fist, and Withers, 2024).

This rule set does not restrict what can be built or trained, nor how models are deployed. It establishes visibility, a verifiable record that governance-relevant compute has occurred, and single-instance integrity for the most capable systems. As the governance infrastructure matures and trust is established, the rule set can be expanded through the multilateral process described in Section V, with each expansion subject to the civil liberties safeguards discussed in Section VII.

No governance mechanism is invulnerable. The right question is not whether the system is invulnerable, but whether it meaningfully raises the cost and complexity of evasion compared to the status quo of no hardware-layer governance at all. Door locks do not prevent all burglaries, but they are a necessary component of any security architecture. Hardware governance raises the floor of enforceable accountability. Nobody claims it eliminates all circumvention.

**V. The Governance Question: Who Sets the Rules?**

**A. The Adoption Pathway Problem**

The primary limiting factor for hardware-layer governance is not technical feasibility but adoption. Securing the participation of chip manufacturers such as Nvidia, TSMC, Intel, and AMD before such mechanisms become regulatory mandates is the hard part. The Future of Life Institute's recommendations for the U.S. AI Action Plan in 2025 proposed, among other measures, requiring continuous affirmative licensing for chips powering a defined category of frontier AI models, with tamper-resistant security modules to prevent circumvention, and geolocation and geofencing capabilities to detect unauthorized deployment abroad (Van Beek, 2025). Whether industry will cooperate voluntarily or await legislative mandates remains to be seen.

The most immediate objection is why leading AI nations would accept embedding verification into their chips. The answer lies in self-interest: the alternative is not the absence of regulation, but fragmentation. In the United States alone, 42 states introduced AI bills in 2025, with divergent approaches ranging from regulatory sandboxes to high-risk frameworks inspired by the EU AI Act, with fewer than 10% of tracked bills actually enacted, revealing both legislative urgency and a lack of coordination (Gluck, Do

& Rice, 2025); internationally, jurisdictions define "AI" in inconsistent ways (White & Case, 2025). When divergences become too large, companies tend to push for harmonization (Lancieri, Edelson & Bechtold, 2025), a logic comparable to that which led nearly 200 nations to adopt International Civil Aviation Organization (ICAO) standards. For chip manufacturers, a multilateral standard is preferable to multiple compliance lines; for nations already employing export controls, hardware governance offers a more sophisticated mechanism than blunt embargo.

This is not an entirely hypothetical scenario. The current U.S. export control regime already operates through a logic of conditional access. The Validated End User (VEU) program, established in 2007, allows pre-approved entities in designated jurisdictions, initially China, to receive designated dual-use items under a general authorization rather than individual export licenses, contingent on compliance monitoring and end-use commitments (Bureau of Industry and Security, 2007; Sutter, 2025).

The now-rescinded AI Diffusion Rule of January 2025 extended this logic through a tiered global licensing framework that included per-company, per-country cumulative computing power caps, a license review policy with a presumption of approval for qualifying destinations, and a bifurcated Data Center VEU program enabling pre-approved entities to receive advanced computing ICs under streamlined authorization (Bureau of Industry and Security, 2025; Sutter, 2025).

Although that rule was withdrawn in May 2025, it signaled an institutional willingness to condition market access on verifiable compliance structures. The August 2025 arrangement between the U.S. government and Nvidia and AMD, which allowed sales of the H20 and MI308 chips to China in exchange for 15% of sale revenues directed to the U.S. government, illustrates an emerging, if legally contested, willingness to use commercial mechanisms as governance tools (Dou & Moon, 2025; Sutter, 2025).

A safety-conditioned export framework would extend this logic on firmer institutional ground. If chip manufacturers embed auditable, tamper-resistant governance modules, the resulting hardware becomes a verifiable compliance point regardless of its destination. The commercial incentive aligns with the governance objective: manufacturers gain expanded market access; governments gain verifiable oversight infrastructure; and even adversarial actors deploying such hardware would operate under constraints embedded in the physical substrate.

For nations in the Global South, the pathway from commercial incentive to operational deployment requires additional institutional scaffolding. Four components would form the basis of such a deployment blueprint:

(1) Funded pilot programs, financed through multilateral development banks, bilateral technology cooperation agreements, or dedicated AI governance funds, could subsidize the initial adoption of governed chips in public-sector applications such as national health systems, agricultural extension services, and educational platforms, where the governance benefits are most visible and the political costs of adoption are lowest.

(2) Regional attestation authorities, established under existing continental or subcontinental institutions such as the African Union, ASEAN, or Mercosur, would ensure that compliance verification is not only made by Global North actors; regional bodies would conduct attestation, handle disputes, and adapt standards to local regulatory contexts, preserving sovereignty over the verification process itself.

(3) Compliance consortia, modeled on the regional Internet registries that distribute IP address space, would pool technical expertise and infrastructure, enabling smaller nations to participate in the governance framework without each independently building the full stack of verification capabilities. (4) Sovereign evaluation and monitoring capacity would require targeted investment in national institutions capable of conducting red-team evaluations, auditing attestation logs, and assessing whether deployed models operate within certified parameters.

Sastry et al. (2024) note that "longer-term capacity-building programs, combining technical and financial assistance, could increase domestic capacity to build and operate AI compute infrastructure in the Global South," adding that "multistate collaborations" to "construct large-scale AI compute for use by the Global South" that "could also spread risk and reap benefits from scale". Without this institutional scaffolding, hardware-layer governance risks reproducing the very asymmetry it aims to correct, embedding verification infrastructure that the Global South can use but cannot govern.

## B. The IAEA Model: Multilateral Oversight for Dual-Use Technology

Nuclear energy faced a structural dilemma similar to the one AI faces today: a dual-use technology, with global reach and catastrophic risk potential, requiring forms of governance that no single nation can provide unilaterally. The International Atomic Energy Agency, established in 1957, had its role strengthened by the Nuclear Non-Proliferation Treaty of 1968, which provided the institutional architecture of the regime: international inspections, multilateral standard-setting, and a framework that distributes oversight beyond any single national actor.

Numerous scholars and practitioners have proposed IAEA-like institutions for AI governance. Maas and Villalobos (2023) cataloged seven distinct institutional models proposed in the literature for international AI governance, including enforcement of standards or restrictions, coordination of policy and regulation, and scientific consensus-building, with the IAEA as one of the most recurrent examples in the enforcement model. Klein and Patrick (2024) explicitly examined the NPT's conditional access regime as a model for AI, noting that the treaty rests on a core bargain whereby non-nuclear-weapon states agree not to acquire nuclear weapons in exchange for nuclear-weapon states' commitment to pursue disarmament and to share access to peaceful nuclear technology, conditioned on compliance with IAEA safeguards.

More recently, Robinson (2025) proposed an International Artificial Intelligence Agency (IAIA) convened by the United Nations, arguing that a new "Grand Bargain" for global AI collaboration is essential, with immediate negotiations between the United States and China to establish an initial IAIA by 2028.

The analysis by Wasil et al. (2024) of international security agreements for dual-use technologies offers important caveats. Although institutions such as the IAEA and the OPCW provide verification functions that the international community has deemed indispensable, enforcement of international agreements remains challenging during geopolitical tensions, as demonstrated by cases of non-compliance. Any AI governance institution would need to account for these limitations, particularly the speed of AI development relative to the pace of international diplomacy, and the risk of regulatory capture by industry actors (Roberts et al., 2024).

A multilateral body combining FlexHEG-style hardware verification with IAEA-style oversight could establish safety standards through multilateral negotiation, certify compliance through on-chip verification, and conduct audits when compliance is contested, addressing sovereignty concerns through privacy-preserving local verification. To move this proposal from analogy to operational specification, three design parameters require elaboration: what the verification regime would attest, what would trigger enhanced scrutiny, and what the graduated response framework would entail.

The IAEA's safeguards system is centered on nuclear material accountancy, quantifying and tracking all nuclear material within a state's territory, complemented by containment and surveillance measures (seals, cameras, and tamper-evident mechanisms), routine and special inspections conducted on both scheduled and short-notice bases, and, where the Additional Protocol is in force, expanded measures such as design information verification and environmental sampling (IAEA, 2022). Routine inspections are not primarily triggered by anomalies; they are programmed activities whose frequency and intensity are determined by facility type, material inventory, and throughput, forming the backbone of the verification regime. Special inspections, by contrast, may be initiated when information available to the Agency, including through routine inspections, is deemed insufficient to fulfill its responsibilities under safeguards agreements.

An analogous regime for compute governance would attest to a defined set of parameters: aggregate training compute consumed (measured in floating-point operations), model architecture class and parameter count, completion of pre-deployment safety evaluations (including red-team assessments against a standardized threat taxonomy), and declared deployment constraints such as geographic scope and use-case limitations. Routine verification would operate through automated attestation logs generated by on-chip mechanisms (see Section IV), analogous to the IAEA's facility operating records, supplemented by scheduled compliance reviews at defined intervals. Enhanced scrutiny, the functional equivalent of special inspections, would be triggered by anomalous compute patterns inconsistent with declared workloads, credible third-party reports of non-compliant deployment, failure to submit scheduled attestation reports, or training runs exceeding a multilaterally agreed compute threshold.

The graduated response framework would follow an escalation logic informed by, though not identical to, the IAEA's enforcement architecture. Initial non-compliance would result in mandatory remediation and enhanced monitoring. Sustained non-compliance would trigger certification suspension, restricting interoperability with governed infrastructure. Serious or repeated violations would escalate to referral to a multilateral enforcement body, mirroring the IAEA Board of Governors' authority to report non-compliance to the UN Security Council and General Assembly, and could prompt coordinated responses by participating states, including export control measures and loss of access to governed compute pools. It should be noted that export controls are instruments applied by states and multilateral arrangements such as the Nuclear Suppliers Group, not direct enforcement tools of the IAEA itself; accordingly, the compute governance analogue would similarly depend on coordinated state action beyond the governing body's own mandate.

Privacy-preserving verification would build on cryptographic attestation techniques, aiming to demonstrate that a computation satisfies declared constraints without revealing model weights, training data, or proprietary architecture, extending the zero-knowledge and trusted execution mechanisms discussed in Section IV. While such mechanisms are technically plausible, their application to the verification of aggregate training compute and architectural properties at scale remains an active area of research rather than a settled engineering problem.

**VI. Reframing the Debate: Open Does Not Mean Ungoverned**

The current policy debate fails by treating "openness" as a final regulatory attribute, when it is merely a distribution property. What matters for governance is the combination of distribution, modifiability, execution environment, audit mechanisms, and institutional oversight authority. A model can be open and still operate under verifiable constraints; a closed model can concentrate significant risks if its use remains opaque. Software distribution controls have limited effectiveness in a world of permissionless digital copying. Hardware, by contrast, is physical, and physical substrates can be governed with enforcement mechanisms that software lacks.

The second key update of the International AI Safety Report, published in late 2025, documented that the number of companies publishing Frontier AI Safety Frameworks more than doubled during the year, while also finding that sophisticated attackers frequently manage to circumvent current defenses (Bengio et al., 2025b). This finding reinforces the case, developed in Section IV, for hardware-layer governance as one necessary component within a defense-in-depth architecture.

The Recommendation of the Council on Artificial Intelligence (OECD AI Principles), updated in 2024 to address general-purpose and generative AI, now explicitly calls for mechanisms to override, repair, and/or safely decommission AI systems that risk causing undue harm (OECD, 2024). The OECD's call for mechanisms to override, repair, and safely decommission AI systems can be read as consistent with hardware-layer governance, though the recommendation does not prescribe a specific technical mechanism. The EU AI Act's GPAI Code of Practice, finalized in 2025, represents a step toward operationalizing these principles, but its reliance on voluntary compliance and compute-based thresholds leaves structural gaps that hardware governance could fill.

Alongside hardware and institutional mechanisms, a complementary governance layer operates at the level of legal liability. The EU's revised Product Liability Directive (Directive (EU) 2024/2853), approved by the European Parliament in March 2024, explicitly classifies software as a product subject to no-fault product liability when supplied in the course of a commercial activity. Under this framework, entities that place AI-enabled products on the market, affix their name or trademark to them, or substantially modify them are treated as manufacturers bearing liability for defective products (Förster, 2024). The Directive exempts free and open-source software developed or supplied outside a commercial activity, but applies fully when open-weight models are distributed as part of a commercial offering or integrated into revenue-generating products (Directive (EU) 2024/2853, Article 2(2); see also Förster, 2024).

This creates a partial but meaningful governance lever. Commercial distributors of open-weight models, cloud providers, platform hosts, and companies offering fine-tuning services, would face legal liability for foreseeable harms arising from defective safety features or the absence of adequate safeguards. Wong's (2025) policy analysis for the R Street Institute similarly proposed risk-tiered liability shields for open-source AI, under which developers of lower-risk models, such as educational tools, would benefit from broad liability shields that encourage innovation while limiting legal exposure in cases of third-party misuse (Wong, 2025), implying that higher-risk applications would not receive equivalent protections.

Legal liability complements technical governance. When commercial actors face material consequences for distributing models without adequate safeguards, the economic calculus shifts toward safety investment. Combined with hardware-level verification and institutional oversight, liability mechanisms complete the defense-in-depth architecture by ensuring that governance operates not only at the point of execution, but also at the point of distribution.

**VII. Limits and Objections**

The argument presented here does not claim that open-weight models are intrinsically safe, nor that restrictions are always counterproductive. There are contexts in which access limits, licensing requirements, export controls, and reporting obligations can reduce real risks, especially when associated with extreme capabilities or specific threats. The narrower claim is that this containment gain is conditional: without alternative governance infrastructure, it may be offset by the displacement effects and dependency dynamics documented in Sections II and III.

Hardware governance itself has limits. Rather than eliminating the concentration of the chip supply chain, it repositions it: the potential advantage is transforming the inevitable concentration of manufacturing into a verifiable compliance point, provided it is subjected to multilateralized and auditable institutional arrangements.

That repositioning, however, introduces a distinct category of risk. Hardware-level controls could be co-opted for purposes unrelated to AI safety: surveillance of legitimate research activity, suppression of commercial competitors, or politically motivated denial of computing access. The concern is not hypothetical. Vipra (2024) documents that effective hardware controls would make it possible for governments to monitor and shut down computing activity considered inconvenient, including normal commercial activity under a state that favors certain corporations, and that the function creep characteristic of surveillance technologies makes this trajectory likely rather than speculative. Heim (2024) observes that even among allied nations, a controlled remote off-switch could significantly deter the widespread use of governed chips, raising concerns about dependence and control that would undermine adoption. Sastry et al. (2024) state the problem directly: compute governance "can be used to promote widely shared objectives like safety, but it can also be used to infringe on civil liberties, prop up the powerful, and entrench authoritarian regimes".

The safeguards required to prevent this outcome are institutional and technical in combination. First, activation of enforcement mechanisms, particularly remote throttling, license revocation, and geofencing, should require multilateral authorization rather than unilateral decision. Sastry et al. (2024) propose multisignature cryptographic protocols in which processing instructions execute only when cryptographically signed by multiple parties, drawing an explicit parallel with the permissive action links (PALs) used in nuclear weapons governance, where multiple authorized individuals must concur before activation. The institutional analogue is that no single government, no single vendor, and no single regulatory body should hold sole authority over the enforcement layer. Second, every governance action, every license issued, every attestation verified, every enforcement event triggered, should be recorded in tamper-evident transparency logs subject to independent audit. Third, governance mandates should incorporate sunset clauses requiring periodic re-authorization, ensuring that rules adopted for one threat environment are revisited as conditions change. Fourth, civil society organizations should hold formal seats in the oversight architecture, with access to audit logs and the standing to challenge governance actions before independent review bodies.

The competition and interoperability dimension is equally consequential. If hardware governance mechanisms are proprietary to a single vendor or a single national jurisdiction, they risk entrenching market incumbents and creating dependency relationships that replicate the very power asymmetries the governance architecture aims to constrain. Open standards for attestation protocols, verification interfaces, and interoperability requirements across hardware vendors are necessary conditions for a governance architecture that distributes power rather than concentrating it.

However, open standards for protocols need not entail open-sourcing the internal implementations of governance processors. The entities that invest in designing and engineering guarantee processors are entitled to intellectual property protection over their implementations, provided those implementations comply with the open attestation standard. A patent-based framework, analogous to the pharmaceutical model under the TRIPS Agreement, would allow a defined period of exclusivity for governance hardware implementations while ensuring that, upon expiration or through licensing, compatible modules can be independently developed by other manufacturers.

The governing principle is that the verification interface must be open so that no single actor controls the compliance layer, while the engineering behind each implementation may remain protected.

Similarly, a multilateral institutional architecture inspired by the IAEA would not automatically replicate the conditions of the nuclear regime, given the differences between fissile materials, replicable software, and far faster innovation cycles in AI. The utility of the analogy is therefore functional and institutional, not literal.

Historical experience with restrictions on dual-use technologies corroborates this distinction. Export control regimes for physical hardware, such as the Wassenaar Arrangement, demonstrate reasonable effectiveness when the controlled object is material and traceable. This effectiveness diminishes when the object is replicable software, as demonstrated by the Crypto Wars of the 1990s: the attempt to restrict the export of strong encryption failed in the face of code's copiable nature (Kehl et al., 2015). Open-weight models, whose weights are instantly distributable digital files, resemble encryption more than centrifuges, reinforcing the need to shift governance to the hardware layer, where the materiality of the substrate permits verifiable intervention points.

The analogy highlights a shared regulatory characteristic: both circulate as copiable digital artifacts, which reduces the effectiveness of purely distributive containment strategies.

**VIII. Conclusion**

Closing access to open-weight models, on its own, displaces the security problem to less visible and less governable spaces rather than resolving it. The evidence reviewed in this paper, including empirical indicators of risk displacement from quantized model proliferation and cross-border API substitution, suggests that restrictions unaccompanied by alternative governance mechanisms create conditions favorable to three recurring effects: proliferation of poorly governed Shadow AI, deepening of the Global South's structural dependency on concentrated infrastructure, and migration of demand to less transparent ecosystems. The threat model taxonomy presented in Section III demonstrates that these risks distribute across hardware, software, institutional, and liability layers, reinforcing the case for defense-in-depth rather than reliance on any single control point.

There is still time to build this kind of arrangement, but less of it as capabilities diffuse and open-weight models become easier to run, adapt, and redistribute outside traditional tracking mechanisms. Under these conditions, regulatory inertia favors more opaque and unequal forms of proliferation.

To move beyond diagnosis, this paper identifies six concrete policy priorities:

(1) Governments and multilateral bodies should fund pilot programs for hardware-enabled governance across the emerging landscape of verifiable compute mechanisms, from chip-level attestation (FlexHEG) and trusted execution environments to software-layer weight protection and confidential computing, beginning with voluntary adoption on next-generation chips, paired with public audits to build institutional trust.

(2) Multilateral negotiations should begin on an institutional framework for AI compute oversight, drawing on the IAEA's graduated verification architecture, routine attestation logs analogous to nuclear material accountancy, enhanced scrutiny triggered by anomalous patterns or non-compliance, and expanded measures for high-capability systems, while adapting to the speed and replicability of digital technologies.

(3) Regulatory frameworks should adopt defense-in-depth principles, layering software-level safeguards (including model-release norms, pre-release safety evaluations, and provenance tracking), hardware-level verification, and institutional oversight rather than relying on any single mechanism, following the logic that Bengio has articulated for AI safety governance: no individual mitigation tool is sufficient, and what is needed is redundancy across multiple governance layers (Bengio, 2024).

(4) Liability regimes should be updated to ensure that commercial distributors of open-weight models bear proportionate legal responsibility for foreseeable harms, drawing on the logic of the EU's revised Product Liability Directive (Directive (EU) 2024/2853), which already classifies software, including

AI systems, as a product subject to no-fault liability when supplied in the course of a commercial activity.

(5) Any hardware-layer governance architecture must incorporate explicit civil liberties safeguards, including multilateral quorum requirements for activating constraints, transparency logs of all governance actions, independent civil society audits, and sunset clauses requiring periodic re-authorization, to prevent the co-option of chip-level controls for censorship or political repression, particularly in jurisdictions with weak rule-of-law protections.

(6) Export control regimes should explore incentive-based mechanisms, such as reduced restrictions for chip manufacturers that implement certified safety features, converting commercial interest into a compliance lever rather than treating regulation and market access as inherently opposed, while funded programs should support governed-chip deployment in the Global South through regional attestation authorities and capacity-building consortia.

The policy challenge is determining at what point, and under what arrangements, the governance infrastructure can be built before its absence imposes higher costs.